\documentclass[12pt]{article}
\usepackage{epsfig}
\usepackage{epsf}
\usepackage{amssymb}
\newcommand{\ba}{\begin{array}}
\newcommand{\ea}{\end{array}}
\newcommand{\bd}{\begin{displaymath}}
\newcommand{\ed}{\end{displaymath}}
\newcommand{\be}{\begin{equation}}
\newcommand{\ee}{\end{equation}}
\newcommand{\bea}{\begin{eqnarray}}
\newcommand{\eea}{\end{eqnarray}}

\def\e{\epsilon}

\def\th13 {\theta_{13}}

\def\bmu {B^{(1,1)}_{\mu}}
\def\wmu {W^{(1,1)}_{3 \mu}}

\begin{document}
\thispagestyle{empty}
\begin{flushright}

\end{flushright}
\vskip 15pt

\begin{center}
{\Large {\bf Probing two Universal Extra Dimensions at International 
Linear Collider}}
\renewcommand{\thefootnote}{\alph{footnote}}

\hspace*{\fill}

\hspace*{\fill}

\large{\bf{
Kirtiman Ghosh \footnote{E-mail address:
kirtiman.ghosh@saha.ac.in},
 Anindya Datta \footnote{
E-mail address: adphys@caluniv.ac.in}
}}\\

\small {\em Department of Physics, University of Calcutta,\\
92, A. P. C. Road, Kolkata 700009, India }
\\

\vskip 40pt

{\bf ABSTRACT}

\vskip 0.5cm

\end{center}

We discuss collider signatures of $(1,1)$-th Kaluza-Klein (KK) mode
vector bosons in the framework of {\em two universal extra dimension
model}, at a future $e^{+}e^{-}$ collider. Production of $\bmu$ and
$\wmu$, the $(1,1)$-th KK mode vector bosons, are considered in
association with a hard photon. Without caring about the decay
products of $\bmu$ or $\wmu$, one can measure the masses of these
particles just by looking at the photon energy distribution. Once
produced $\bmu$ ($\wmu$) dominantly decays to a pair of jets or to a
pair of top quarks. Thus we look for a pair of jets or a pair of top
quarks in association with a photon. Upto the kinematic limit (with 
$e^+ e^-$ center-of-mass energies of 0.5 TeV and 1 TeV) of the
collider, signals from the $\bmu$ production and decay in both the
above mentioned channels are greater than the $5\sigma$ fluctuation of
the Standard Model background with $500 ~fb^{-1}$ integrated
luminosity. However, the number of events from $\wmu$ production and
decay is smaller and its detection prospect is not very good.
\vskip 30pt

\section{Introduction}

Recently lots of attention have been paid to the models of
fundamental interactions with one or more extra space like dimensions
\cite{add,rs}. There is a class of such interesting models where all the
Standard Model (SM) fields can access these extra space-like
dimensions along with the (3+1) dimensional Minkowski space
time. These are collectively called the Universal Extra Dimensional
(UED) models \cite{ued}.

A particular variant of the UED model where all the SM fields
propagate in $(5+1)$ dimensional space time, namely the {\em two
Universal Extra Dimension} (2UED) Model has some attractive
features. 2UED model can naturally explain the long life time for
proton decay \cite{dobrescu} and more interestingly it predicts that the
number of fermion generations should be an integral multiple of three
\cite{dobrescu1}.


As the name suggests, in 2UED, all the SM fields can propagate
universally in the six-dimensional (6D) space-time. Four dimensional
(4D) space time coordinates $x^{\mu}$ ($\mu=0,1,2,3$) form the usual
Minkowski space. Two extra spacial dimensions with coordinates $x^4$ and
$x^5$ are flat and are compactified with $0\le~x^4,~x^5~\le
L$. Toroidal compactification of the extra dimensions, leads to 4D
fermions that are vector-like with respect to any gauge
symmetry. Alternatively, one needs to identify two pairs of adjacent
sides of the square. This compactification mechanism automatically
leaves at most a single 4D fermion of definite chirality as the zero
mode of any chiral 6D fermion \cite{dobrescu2}.

The requirements of anomaly cancellation and fermion mass generation
force the weak-doublet fermions to have opposite {\em 6D chiralities}
with respect to the weak-singlet fermions. So the quarks of one
generation are given by $Q_+~\equiv~(U_+,D_+),~U_-,~D_-$. The 6D
doublet quarks and leptons decompose into Kaluza-Klein (KK) towers of
heavy vector-like 4D fermion doublets with left-handed zero mode
doublets. Similarly each 6D singlet quark and lepton decompose into
the KK-towers of heavy 4D vector-like singlet fermions along with zero
mode right-handed singlets. These zero mode fields are identified with
the SM fermions. In 6D, each of the gauge fields, has six
components. Upon compactification, they give rise to 
towers of physical 4D massive spin-1 fields and a tower of spinless
adjoints. In a previous work \cite{KGAD1} we have discussed the
phenomenology of these spinless adjoints in some details. 
In this letter, we will be interested in a particular member 
of the KK-towers of hypercharge gauge boson $B_\mu$ and $SU(2)$ 
gauge boson $W^3_\mu$.

We would like to investigate the production of $B_\mu$ and $W^3_\mu$
in association with a {\em hard} photon at a future $e^+ e^-$ linear
collider. Somewhat similar things have been discussed in
Ref. \cite{kong}. Authors in Ref. \cite{kong}, have considered the
production of $B_\mu$ is association with a photon. However, they
demand that the photon is undetectable and is lost along the beam
pipe.  This implies that the identification and mass determination of
$B_\mu$, crucially depend on jet (coming from the decay of $B_\mu$)
reconstruction and jet energy measurement.  In contrast, we look for a
final state consisting of a {\em hard} photon and the decay products
(which may or may not be detectable always) of $B_\mu$ and
$W^3_\mu$. In some sense our method is complementary to that used in
Ref. \cite{kong}. The advantages of tagging the photon, will be illuminated
in the next section.

The tree-level masses for $(j,k)$\footnote{Each member of a KK-tower
is specified by a pair of integers, called the KK-numbers.}-th KK-mode
particles are given by $\sqrt{M_{j,k}^2~+~m_{0}^{2}}$, where
$M_{j,k}=\sqrt{j^2+k^2}/R$. The radius of compactification, $R$,
is related to the size of the extra dimensions, $L$ via the relation $L =
\pi R$. $m_0$ is the mass of the corresponding zero mode
particle. As a result, the tree-level masses are approximately
degenerate. This degeneracy is lifted by radiative corrections.

Conservation of momentum (along the extra dimensions) in the full
theory, implies KK number conservation in the effective 4D
theory. SM-like interactions in the 6D, (called the {\em bulk
interactions}) give rise to the the {\em KK-number conserving} 
as well as {\em KK-parity conserving} interactions, 
in 4D effective theory after compactification. 
However, one can generate KK number violating (KK parity 
conserving) operators at one loop level, starting from the
bulk interactions. Structure of the theory demands that
these operators can only be on $(0,0,),~(0,L)$ and $(L,L)$ points of
the {\em chiral square}. In this letter, we will exploit one such
KK-number violating coupling to find a characteristic signature of
2UED model at an $e^+ e^-$ collider.  Namely, we will discuss the
collider signatures of $\bmu$ and $\wmu$, the $(1,1)$-th KK excitations of
the $U(1)$ and neutral $SU(2)$ gauge bosons. $\bmu$
($\wmu$) couples to an electron-positron pair via KK-number violating
coupling \cite{dobrescu3}:
 
\be
{\cal L} = \left [ \bar e\;\left(c_{L}^{V} P_L + c_{R}^{V} P_R \right )\gamma^{\mu}e \right ] V_{\mu}^{(1,1)}.
\label{coup}
\ee
Where, 
\bea
c_{L}^{B}&=&  \frac{g^\prime g^2}{16 \pi^2} \left(\frac{9}{8}+\frac
{91\;g^{\prime 2}}{24\;g^{2}} 
\right )\;\ln \frac{M_s^2}{M_{j,k}^2}, \nonumber \\
c_{R}^{B}&=& \frac{g^{\prime 3}}{16 \pi^2}  
\left(\frac{59}{6}\right ) \;\ln \frac{M_s^2}{M_{j,k}^2}, \nonumber \\
c_{L}^{W^{3}}&=& \frac{g^3}{16 \pi^2}\left(-\frac{11}{24} +
\frac{3g^{\prime 2}}{8g^{2}}\right ) \;\ln \frac{M_s^2}{M_{j,k}^2},
\nonumber \\ 
c_{R}^{W^{3}}&=& 0.  
\label{clcr}
\eea
  These couplings also have logarithmic dependence on the cutoff scale,
  $M_s$, of the theory. We assume $M_s$ to be 10 times the
  compactification scale $R^{-1}$ following \cite{dobrescu3}.

 Contributions to the KK-number violating operators like
 Eq. (\ref{coup}) might be induced by physics above the cut-off
 scale. We assume that those UV generated localized operators are also
 symmetric under KK parity, so that the stability of the lightest KK
 particle which can be a promising dark matter candidate \cite{darkUED}, is
 ensured. Loop contributions by the physics below cut-off scale $M_s$
 are used to renormalize the localized operators \cite{ponton}.

\section{Signatures at future $e^+e^-$ collider with photon tag}
 Resonance production of $\bmu$, has been investigated in the context
 of Tevatron and LHC in \cite{dobrescu3, dobrescu4} and in the context
 of future $e^+ e^-$ collider in \cite{kong}. However, in this letter,
 we will reconsider the prospects of $\bmu$ (also $\wmu$) production
 and detection at future $e^+e^-$ colliders, exploiting the KK-number
 violating couplings defined in Eq. (\ref{coup}).

 There is a disadvantage of $e^+e^-$ collision. Unless the mass of the
 particle, we want to produce, matches exactly with the $e^+e^-$
 center-of-mass energy, resonance production cross-section is miniscule. This
 compels us to consider the $\bmu$ ($\wmu$) production in association
 with a photon ($\e^+e^-\to \gamma \bmu, \gamma \wmu$). This particular 
production mechanism has many interesting
 consequences.  First of all, just measuring the photon energy one can
 have the knowledge of the mass of $\bmu$, without caring about the
 decay products of $\bmu$. Moreover, we will also notice that, the
 production cross-section grows with mass of $\bmu$
 ($W_{3\mu}^{(1,1)}$).

\begin{figure}[t]  
\centerline
{\rotatebox{270}{\epsfxsize=8.cm\epsfysize=8cm\epsfbox{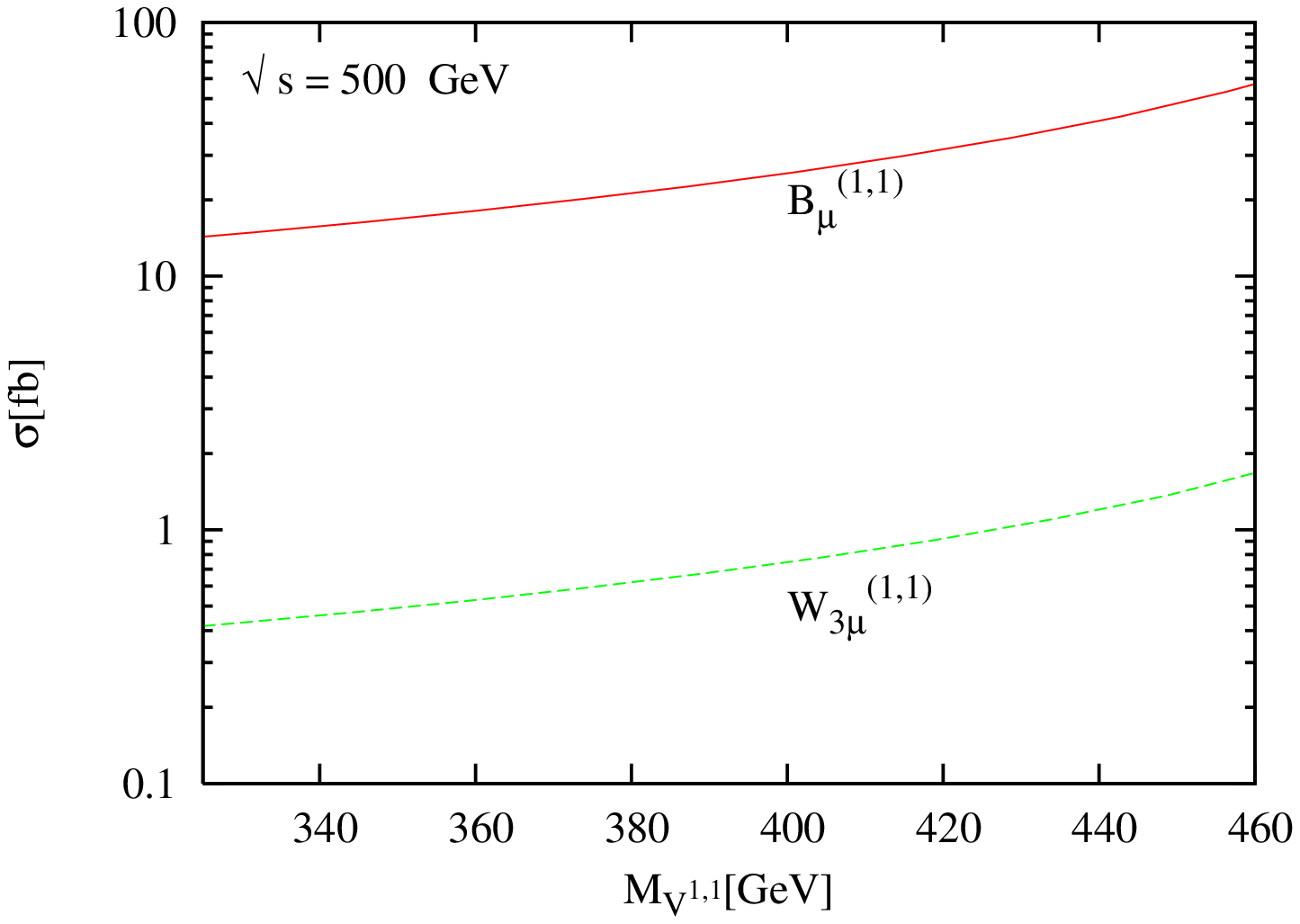}}}
\centerline
{\rotatebox{270}{\epsfxsize=8.cm\epsfysize=8cm\epsfbox{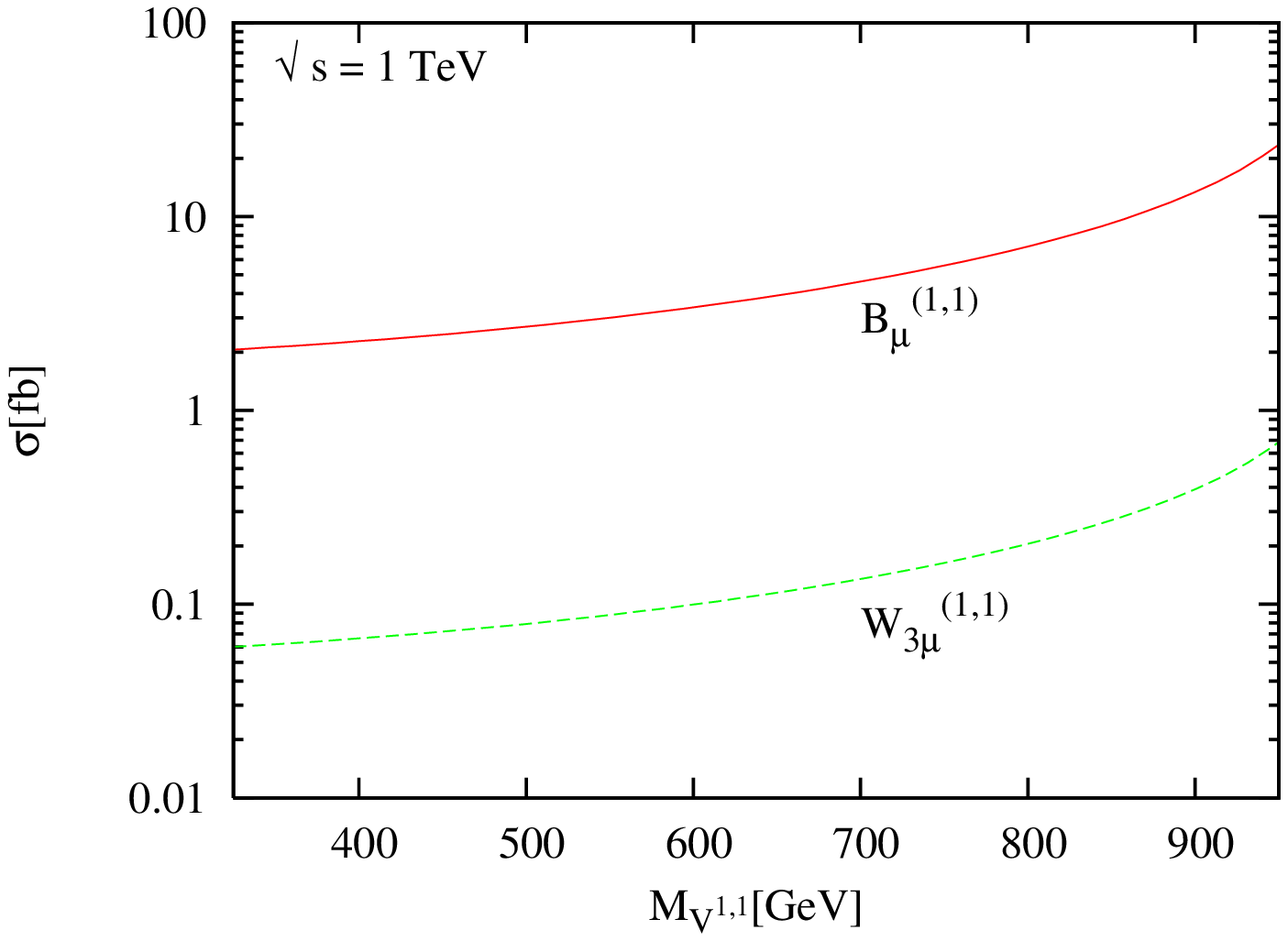}}}

\caption{Cross-sections (fb) of $e^+ e^- \rightarrow \gamma \;
  \bmu({\rm solid line}), ~\gamma\;\wmu ({\rm dashed line})$
for $e^+ e^-$ center-of-mass energies 0.5, 1 TeV
respectively.}
\label{cross-sec}
\end{figure}

 $\bmu$ and $\wmu$ production in association with a photon takes place
 in $e^+ e^-$ collision, via t(u) channel.  Spin averaged matrix
 element squared at the LO is given by :

\be
\overline {\sum \vert {\cal M} \vert^2} = 4\pi \alpha_{em}\;(c_{L}^{V^{2}} + c_{R}^{V^{2}})
\,\left( \frac{u}{t} + \frac{t}{u} + \frac{2m_{V}^{2}s}{ut}\right),
\label{cross}
\ee

\noindent 
s, t, u are the usual Mandelstam variables, and $c^{V}_{L},~c^{V}_{R}$ are
defined in Eq. (\ref{clcr}). The numerical values of the cross-sections are
presented in Fig. \ref{cross-sec} against the masses of $\bmu$ and
$\wmu$ for two different values of $e^+ e^-$ center-of-mass energies.
Fig. \ref{cross-sec} shows a very interesting variation of
cross-section . In spite of the fact that, the couplings in Eq. (\ref{coup}) 
do not
increase with the masses or $R^{-1}$, 
the cross-section increases when the mass of $V_{\mu}^{(1,1)}$
approaches closer to the center-of-mass energy, which is fixed for a
particular collider. This, in fact, is a more general phenomena not
specific to the 2-UED model. The probability of the photon emission
from one of the initial $e^-$ or $e^+$, increases with the diminishing
photon energy. One can easily check that for a fixed center-of-mass
energy ($\sqrt s$) of the collider, photon energy $E_\gamma$ is given
by: $\frac{s - m^2}{2\sqrt{s}}$. Thus a KK gauge boson mass closer to
the center-of-mass energy reduces the photon energy which in turn
increases the cross-section. Similar effects can take place in the cases
of single production of sneutrinos \cite{skrai1} (in association with
a photon) via lepton number violating couplings; graviton production
in ADD or RS model (in association with a photon)
\cite{skrai2}.

The increase of cross-section with mass can also be very easily understood
by looking at Eq. (\ref{cross}). Both, $u$ and $t$ are proportional to
the photon energy $E_{\gamma}$. An
increasing $\bmu$ or $\wmu$ mass would mean (for a fixed $e^+ e^-$
center-of-mass energy) a diminishing u and t. This in turn enhances
the cross-section with mass.
  
Rate of $\bmu$ production is always an order of magnitude higher than the
rate of $\wmu$ production over the mass range upto the kinematic
limit. $\wmu$ couples only to the left-handed electrons via the
$SU(2)$ gauge coupling. On the other hand, $\bmu$ couples to both left-
and the right-handed electrons (see Eq. (\ref{clcr})). Moreover, a
partial cancellation between two terms in the expression of
$c_L^{W^3}$ makes the $W_{3\mu}^{(1,1)}$ production cross-section
smaller. The dominance of $\bmu$ 
cross-section over the $\wmu$ can be partially explained from these
couplings.

We can now discuss the signals of $\bmu$ and $\wmu$ production at $e^+
e^-$ collisions. Once produced, $\bmu$ ($\wmu$) dominantly decays to
a pair of light quark jets.
It also decays to a $b \bar b$ or $t \bar t$ pair. We collectively look for two
jets (light or $b-$flavoured) from the decay of $\bmu$ or $\wmu$ and
a nearly mono-energetic photon.  If we look at the energy distribution
of the photons, $\bmu$ and $\wmu$ production would be characterised by
two (mono-energetic) peaks separated by, $\Delta E_{\gamma}
 = \frac{m_{\wmu}^2 - m_{\bmu}^2}{2 \sqrt{s}}$. 
\begin{figure}[h]  
\centerline{
\rotatebox{270}{\epsfxsize=8cm\epsfysize=10cm\epsfbox{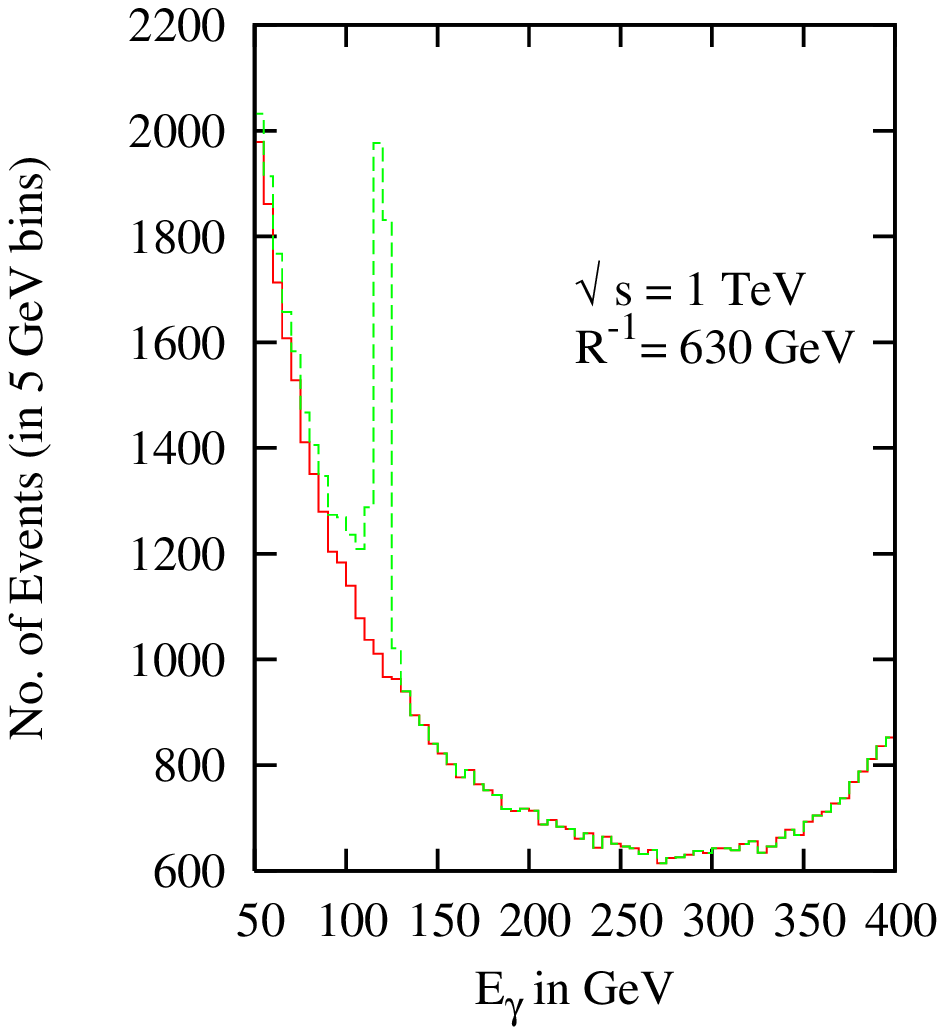}}
}
\caption{Photon energy distribution for $\gamma + 2j$-events for
signal (dashed histogram) and background (solid histogram). The 
monoenergetic (in case of the signal)
photon peak is smeared due to ISR effects and finite detector
resolution. We have used $R^{-1} = 630$ GeV, and $\sqrt{s_{ee}} = 1$ TeV.}
 
\label{energy_dist_1}
\end{figure}

Production of $\bmu$ ($\wmu)$, in association with a photon, is
twofold advantageous. Instead of a fixed center-of-mass energy, now
the effective center-of-mass energy of the collision (which produces
the new physics) can vary over a range thus makes it possible to
produce $\bmu$ and/or $\wmu$ with different masses.  Moreover, by
measuring the energy of the photon, we can determine the masses of
$\bmu$ and $\wmu$ without caring about the decays of these
particles\footnote{ Similar technique has been exploited in
\cite{sakurai} to find the signals of doubly-charged Higgs at an $e^-
e^-$ collider.}.  $\bmu$ or $\wmu$ dominantly decays to a pair of
jets.  One can thus measure the masses of $\bmu$ or $\wmu$, directly
by measuring the jet energies. Authors in Ref. \cite{kong}, have
investigated the production of $\bmu$ in $e^+ e^-$ collision. They
have emphasised on directly measuring the jet energies and
reconstructing the $\bmu$ mass.  This involves, identification and
energy measurement of {\em both} the jets coming from the $\bmu$
decay.  However, photon identification and measurement of its energy in
electromagnetic calorimeter can be done more easily in comparison to 
the same exercise with the jets.

For an ideal detector with infinitely high resolution, the photon
energy distribution is ideally an delta-function at $E_{\gamma} =
\frac{s - m_{V_\mu}^2}{2
\sqrt{s}}$. As a consequence of finite detector resolution and initial state
radiation (ISR) the photon energy distribution is smeared. However, the
effects which smear the $E_\gamma$ peak, cannot change the position of
the peak, enabling us to measure the masses of $\bmu$ or $\wmu$ just
by looking at the position of the peaks in the $E_{\gamma}$
distribution. This method works well, independent of any particular
decay mode of $\bmu$ ($\wmu$). As for example, one can consider the
case of $\bmu$ decaying to $t \bar t$ (branching ratio of $\bmu (\wmu)
\to t
\bar t$ is 30 (15) \%). Final state comprises of missing energy/momentum 
due to the presence of neutrinos if one allows the top quarks to decay
semi-leptonically. In such a situation, reconstructing the $\bmu$ mass
will be difficult. Even when the top quarks decay hadronically, 
we have to be careful about reconstructing the two top quarks out of
the six jets. This would be a challenging task. However, just by
looking at the nearly mono-energetic photon, we ease our task by a
considerable amount.

We have also estimated the SM contribution to the $\gamma + 2j$ final
state. Fig. \ref{energy_dist_1} shows the $E_\gamma$ distributions for signal
(dashed histogram) and background (solid histogram) for an $e^+ e^-$
center-of-mass energy of 1 TeV. We have used $R^{-1} = 630 \rm
~GeV$ for the purpose of illustration in this figure.  ISR effects have
been included in our analysis following the prescription in
Ref. \cite{isr}. To include a realistic detector response, we have smeared 
the photon and jet momenta using a Gaussian smearing \cite{smear}.
The topology of signal and background events are more or less the
same.  As a result, the kinematic cuts defined below are for the
purpose of selection only.

The following selection criteria are applied on signal and
backgrounds:\\ 
$p_T^\gamma > 10 ~\rm GeV$, $p_T^j > 20 ~\rm GeV$, \\
$\vert \eta_\gamma \vert < 2.5$, $\vert \eta_j \vert < 3$, \\ $\Delta R~
(\equiv \sqrt{\Delta \eta ^2 + \Delta \phi^2})$ (between any pair of
photon and jets) $ > 0.7$.\\

\begin{table}[h]
\begin{center}
\begin{tabular}{|c|c|c|c|c|c|c|c|}
\hline 
$e^+e^-$&$R^{-1}$ & \multicolumn{3}{c|}{$B_{\mu}^{(1,1)}$} &
\multicolumn{3}{c|}{$W^{3(1,1)}_{\mu}$} \\\cline{3-8} 
C-o-M&in       & $m_{B_{\mu}^{(1,1)}}$ & Signal & Background &
$m_{W^{3(1,1)}_{\mu}}$ & Signal & Background \\ 
Energy& GeV     & GeV & Event & Event & GeV & Event & Event \\ \hline
&280 & 387.3 & 5900 & 19258 (139)& 433.8 & 253 & 26593 (163) \\
&290 & 401.1 & 6713 & 20368 (143)& 448.7 & 349 & 34031 (184) \\
500&300 & 414.9 & 7701 & 22207 (149)& 463.7 & 520 & 50011 (224) \\
GeV&310 & 428.8 & 9005 & 24814 (158)& 478.7 & - & - \\
&340 & 470.3 & 24296 & 59938 (245)& 523.6 & - & - \\ \hline \hline
&300 & 414.9 & 348 & 2889 (54) & 463.7 & 10 & 2499 (50) \\ 
&400 & 553.3 & 430 & 2038 (45)& 613.8 & 14 & 1932 (44)  \\ 
1&550 & 760.8 & 948 & 2096 (46)& 840.4 & 43 & 2538 (50) \\ 
TeV&630 & 871.4 & 2082 & 3013 (55)& 961.5 & 210 & 8444 (92) \\ 
&690 & 954.4 & 6552 & 7482 (87)& 1052.4 & - & -\\ \hline 
\end{tabular}
\end{center}
\caption{Number of $\gamma + 2j$ signal and SM background events for 
two values of $e^+e^-$ center-of-mass energies assuming 500 $fb^{-1}$ 
integrated luminosity. $1\sigma$ fluctuations of the background events are
also shown in the brackets. The entries marked with a dash, correspond to the
situations when number of events are too small, or $\bmu$ ($\wmu$) production
is kinematically disallowed.
   }
\label{2jet_tabl}
\end{table}

In Table \ref{2jet_tabl}, the total number of signal events in the
bins corresponding to the peak in the photon energy distributions and
its two adjacent bins are presented for different values of $R^{-1}$. We
have used a bin size of 5 GeV. The total number of background events
corresponding to the above three bins are also presented with their
$1\sigma$ fluctuations. It is evident from the table, almost upto the
kinematic limit of the $e^+ e^-$ collision, signal from $\bmu$ is
always greater than the $5\sigma$ fluctuation of the 
background. However, the signal from $\wmu$ is weaker and merely can
surpass the $1\sigma$ fluctuation of the SM background for $\wmu$ masses
closer to the $e^+e^-$ center-of-mass energy. Thus it is not
possible to measure both the peaks over the SM background. This in
turn kills the hope to measure the correlation between the masses and
the cross-sections of the $\wmu$ and $\bmu$ production in 2UED.

Now we will discuss the situation when $\bmu$ or $\wmu$ decays to $t \bar t$.
Final state thus consists of a monoenergetic photon with decay
products coming from the pair of top quarks. Instead of incorporating
the detailed decay and reconstruction of top quarks at the detector
level, we have multiplied our cross-sections by top reconstruction
efficiency (0.55) in 6-jet and 4-jet plus 1-lepton channel
\cite{top_reconstruct} in our analysis. In Table \ref{top_tabl}, the numbers 
of $\gamma + 2t$ events corresponding to the bin (and its two adjacent
bins) for which $E_\gamma$ distributions shows the characteristic peak,
are presented for signal and background. Number of $2t$ events
from $\bmu$ are smaller with respect to the $2j$ events, due to
smaller branching ratio and top-reconstruction efficiency. Number of
background events are also smaller in $2t$ channel compared to the
$2j$ channel. 

Number of events from the $\bmu$ production and decay (either in $2j$
or $2t$ mode) are always well above the $5\sigma$ fluctuations of the
SM background.  This opens up a possibility, to measure cleanly the
relative strengths of the signals from $\bmu$ decaying into $2j$ and $2t$
channels\footnote{Modulo the detection efficiencies in both these
channels, which could be determined beforehand from simulation and
experimental data.}. Consequently one can determine the {\em ratios of
the decay widths} of $\bmu$ into $jj$ mode and $t \bar t$ mode. This
ratio is {\em not sensitive} to the cut-off scale $M_s$ unlike the
cross-sections.  Apart from the coupling constants, the ratio depends
only on $\bmu$ mass (not on other parameters like $R$ or $M_s$).  Mass
of $\bmu$ also can be measured independently from the peak position of
$E_\gamma$ distribution.  Using this value of experimentally measured
mass, one can calculate the ratio as in the 2UED model. Finally, this
theoretical number can be compared with the experimentally measured
ratio of decay widths.

\begin{table}[h]
\begin{center}
\begin{tabular}{|c|c|c|c|c|c|c|c|}
\hline 
$e^+e^-$&$R^{-1}$ & \multicolumn{3}{c|}{$B_{\mu}^{(1,1)}$} &
\multicolumn{3}{c|}{$W^{3(1,1)}_{\mu}$} \\ \cline{3-8} 
C-o-M&in       & $m_{B_{\mu}^{(1,1)}}$ & Signal & Background &
$m_{W^{3(1,1)}_{\mu}}$ & Signal & Background \\ 
 Energy&GeV     & GeV & Event & Event & GeV & Event & Event \\ \hline
&250 & 345.8 & - & - & 389.1 & 8 & 484 (22)\\ 
&280 & 387.3 & 519 & 484 (22)& 433.8 & 18 & 774 (28)\\
500&295 & 408.1 & 776 & 506 (23)& 456.2 & 30 & 1305 (36)\\ 
GeV&310 & 428.8 & 1115 & 711 (27)& 478.7 & 46 & 1673 (41)\\ 
&340 & 470.3 & 3586 & 2248 (48)& 523.6 & - & - \\ \hline \hline
&300 & 414.9 & 40 & 63 (8)& 463.7 & - & - \\ 
1&400 & 553.3 & 76 & 77(9) & 613.8 & - & - \\ 
TeV&550 & 760.8 & 189 & 126 (11)& 840.4 & 5 & 178 (13)\\ 
&630 & 871.4 & 461 & 245 (16) & 961.5 & 25 & 747 (27) \\ 
&690 & 954.4 & 1482 & 654 (26)& 1052.5 & - & -\\ \hline
\end{tabular}
\end{center}
\caption{Number of $\gamma + 2t$ signal and SM background events for 
two values of $e^+e^-$ center-of-mass energies assuming 500 $fb^{-1}$ 
integrated luminosity. $1\sigma$ fluctuations of the background events are
also shown in the brackets.  The entries marked with a dash, correspond to the
situations when number of events are too small, or $\bmu$ ($\wmu$) production
is kinematically disallowed or $\bmu$ ($\wmu$) decay to $t \bar t$ is 
kinematically not possible .
 }
\label{top_tabl}
\end{table}

Number of $\gamma + 2t$ events from $\wmu$ production is again small
and cannot compete with the SM background. For the sake of completeness,
we have presented these numbers also in Table \ref{top_tabl}.
 
\section{Conclusion}

To summarise, we have discussed a possible signature of $\bmu$ and
$\wmu$ production along with a hard photon, in the framework of 2UED
model, at a future $e^+e^-$ collider. Once produced these gauge bosons
decay either to a pair of light quarks or to a pair of top quarks. So
the signatures of these vector bosons are a pair of jets or a pair of
top quarks with a nearly monoenergetic photon. Production of these
$(1,1)$-mode gauge bosons along with a single hard photon is
advantageous. Without caring about the decay products of $\bmu$ and
$\wmu$, one can measure the masses of these particles by measuring the
energy of the photon. Number of signal events from $\bmu$ production
is always greater than the $5\sigma$ fluctuation of the SM background,
for $R^{-1}$ values up to the kinematic limit of the collision. Rate
of $\wmu$ production is small and cannot stand over the SM background
in either $2t$ or $2j$ channel. Thus the measurement of the possible
correlation between the masses of $\bmu$ and $\wmu$ and their signal
strengths is not possible. However, the number of events from $\bmu$
production and decay (both in $\gamma + 2j$ and $\gamma + 2t$ channels)
are large. These enable one to measure the cross-sections in these
channels precisely.  The relative strength of the $\gamma + 2j$ and
$\gamma + 2t$ signals thus can be measured. This ratio of the
cross-sections are equal to the ratio of $\bmu$ decay
widths into $jj$ and $t \bar t$ channels. Interestingly this ratio is 
independent of the cut-off scale of the theory. Thus experimentally
measured ratio can be contrasted with the theoretical predictions from
2UED model.

\noindent
{\bf {Acknowledgments}} KG acknowledges the support from
Council of Scientific and Industrial Research, Govt. of India. AD is
partially supported by Council of Scientific and Industrial Research,
Govt. of India, via a research grant 03(1085)/07/EMR-II. AD is also 
partially supported by DAE-BRNS research grant 2007/37/9/BRNS.\\

\hspace*{\fill}

\end{document}